\documentclass[12pt]{article}
\usepackage{amssymb}
\usepackage{amsmath}
\begin{document}

\title{Variational formulation  for models of shear shallow water flows and  ideal turbulence}
\author{ Sergey L. Gavrilyuk \thanks{Corresponding author. E-mail:
		sergey.gavrilyuk@univ-amu.fr } \ and\ \ Henri Gouin \thanks{   
	  E-Mails:
		henri.gouin@univ-amu.fr; henri.gouin@yahoo.fr} }
	\date{\footnotesize Aix Marseille Univ, CNRS,
		IUSTI, UMR 7343, Marseille, France.}
	\maketitle
	\begin{abstract}
		The shallow water equations without shear effects
		are similar to the  gas dynamics equations with a polytropic
		equation of state. When the shear effects are taken into account, the  equations contain  additional evolution equations mathematically analogous to those of  the  Reynolds stresses in turbulent flows of  compressible fluids when the source terms are  neglected (\textit{ideal turbulence}).    We show that the non-dissipative model of shear shallow water flows and the  model of ideal turbulence   admit  a similar variational formulation  where, in the both cases, the  equations for  the Reynolds stress tensor evolution are considered as  non-holonomic constraints.

	\end{abstract}
	
	
	\section{Introduction}
	The reason for considering  together   simplified models of
	non-dissipative shear  shallow water flows and  ideal turbulence without source terms is   there exists a  mathematical similarity between an  asymptotic model
	of shear flows of long waves over flat bottom and  the homogeneous Reynolds-averaged equations in turbulence \cite{Teshukov_2007,Gavrilyuk_Gouin_2012,Richard_2012,Richard_2013,Berthon_2002, Audebert_2008}. 

	The two models  can be considered as  reversible because they admit  a  conservative energy balance law.  The equation for   Reynolds stress tensor evolution and its analogue for shear shallow water flows cannot be integrated: it  corresponds  to non-honolomic constraints in   analytical mechanics. This observation allows us to formulate  the Hamilton principle for  shear  shallow water flows and for ideal turbulence    with a  non-holonomic constraint governing   the Reynolds stress tensor evolution.
	
The structure of the  paper is as follows.\\  In section 1, we present the both models and establish their equivalence in the dissipationless limit.\\ In section 2, we derive the momentum equation for both models from Hamilton's principle of stationary action by considering the Reynolds equations as non-holonomic constraints.  \\
A conclusion ends the paper.
	\subsection {Shear shallow water flows}
	
	The governing equations of  shear shallow water flows are 
 derived from   free surface Euler equations  as an exact asymptotic model of weakly sheared flows for long waves over a flat bottom \cite{Teshukov_2007} (and \cite{Castro_Lannes_2014}, where is given a generalization of the model):
	
	\begin{equation}
		\left\{
		\begin{array}{l}
			\displaystyle\frac{\partial h}{\partial t}+{\rm div}\left(  h\,\mathbf{U}%
			\right) =0, \\
			\\
			\displaystyle\frac{\partial (h {\mathbf{U}})}{\partial t}+\left[\mathrm{div}\left( h  {\mathbf{U}}\otimes {\mathbf{U}}+\frac{gh^{2}}{2}\,\mathbf{I}+ {\mathbf{R}}\right)\right]^T =0,\\
			\\
		  \displaystyle\frac{d\mathbf{R}}{dt}+\mathbf{R}\,\mathrm{div}\mathbf{U}%
			+\frac{\partial \mathbf{U}}{\partial \mathbf{x}}\ \mathbf{R}+\mathbf{R}%
			\left( \frac{\partial \mathbf{U}}{\partial \mathbf{x}}\right) ^{T}=0.
		\end{array}%
		\right.   \label{Teshukov_system}\\
	\end{equation}\\
	
	\noindent The superscript "$^T$\," denotes the transposition, $\mathbf{I}$ is the identity tensor, $\displaystyle   {d}/{dt}= {\partial }/{\partial t}+\mathbf{U}^{T}\ \nabla\ $   is the material derivative with respect to
	the mean motion, $h(t,\bf x)$ is the fluid depth and $g$ is the gravity acceleration.  The depth average velocity $\mathbf U$  and the Reynolds stress tensor $\mathbf R$  are defined as :
	$$
	\qquad h\,\mathbf{U=}\int_{0}^{h}
	\widetilde{\mathbf{U}}\; dz,\qquad {\mathbf{R}}=\int_{0}^{h}(\widetilde{\mathbf{U}}-\mathbf{U})\otimes (%
	\widetilde{\mathbf{U}}-\mathbf{U})\; dz,
	$$%
where $\widetilde{\mathbf{U}}(t,{\bf {x}}, z)$ is the instantaneous fluid velocity. Variables ${\bf x}=(x,y)^T $ are the horizontal coordinates, $z$ is the vertical coordinate (which is opposed to the direction of   gravity acceleration).  
	Equations are written in 
	the limit of weakly sheared flows for three-dimensional long waves.\\ The reduced Reynolds stress tensor is : 
$${\bf P}=\frac{\bf R}{h}$$ and
System (\ref{Teshukov_system}) admits the energy conservation
	law :
	\begin{eqnarray}
		\label{energy1}
		&&\frac{\partial }{\partial t}\left(h\left( \frac{1}{2}{\left\vert \mathbf{U}\right\vert ^{2}}+ \frac{gh}{2}+\frac{\mathrm{tr}\,\mathbf{P}}{2}\right) \right)\\ &&+\,\mathbf{\mathrm{div}}
		\left( h\mathbf{U}\left( \frac{1}{2}{\left\vert \mathbf{U}\right\vert ^{2}}+\frac{gh}{2}+\frac{\mathrm{tr}\,\mathbf{P}}{2}\right) +\frac{gh^2}{2}\mathbf{U}+h\mathbf{P}\mathbf{U}\right) =0. \notag
	\end{eqnarray}
	 Taking into account the mass conservation law \eqref{Teshukov_system}$_1$, (equation for $h$), we obtain the  equivalent equation for $\bf P$ :
	\begin{equation*}
		\frac{d\mathbf{P}}{dt}+\frac{\partial \mathbf{U}}{\partial
			\mathbf{x}}\
		\mathbf{P}+\mathbf{P}\left( \frac{\partial \mathbf{U}}{\partial \mathbf{x}}%
		\right) ^{T}=0.  \label{basic_equation}
	\end{equation*}%
	.
	\subsection{Ideal turbulence}
	The Reynolds averaging of the turbulent flows for barotropic compressible fluids is (for example refer to \cite{Pope_2005,Wilcox,Mohammadi_1994}) :
	
	\begin{equation}
	\left\{ 
	\begin{array}{l}
	\displaystyle\frac{\partial \overline{\rho}}{\partial t}+(\overline{\rho u_i}),_{i}=0, \\
	\\
		\displaystyle\frac{\partial \overline{\rho u_i}}{\partial t}+\left( \overline{\rho u_i u_j} +\overline{p} \,\delta
	_{ij}\right),_{j}=0, \\
	\\
		\displaystyle\frac{\partial \overline{\rho u_i u_s}}{\partial t}+ \left(\overline{\rho u_i u_k u_s}\right),_k +\overline{p,_i u_s}+\overline{p,_s u_i}=0,
	\end{array}
	\right.
		\label{System0}
	\end{equation}
where $\rho$ is the fluid density, $\mathbf u=[u_1,u_2,u_3]^T$ is the instantaneous velocity field, \textit{comma}  means the derivative with respect to the Eulerian coordinates $x_i$, $i=1,2,3$,  $p=\mathcal{P}(\rho)$ is a given equation of state,  $\delta
	_{ij}$ denotes the Kronecker symbol and repeated indexes mean the summation.\\ For any function $f$  we have denoted by $\overline f$ the Reynolds averaging (for example, time or space averaging). The last equation is obtained by multiplying the momentum equation by $u_s$ and  by averaging this new   equation. 
	
	In the case of compressible fluids, we usually use  \textit{the Favre  averaging velocity} (mass averaging velocity) \cite{Favre_1976} :
	\begin{equation*}
	{\mathbf U}=[U_1,U_2,U_3]^T, \quad U_i=\frac{\overline{\rho u_i}}{\overline{\rho}}.
	\end{equation*}
	In this case, the mass equation yields   : 
	\begin{equation*}
		\overline{\rho}_{t}+(\overline{\rho} U_i),_{i}=0
	\end{equation*}
	Introducing the following definitions of fluctuations of density and pressure :
		\begin{equation*}
\rho=	\overline{\rho}+\rho^\prime, \quad  \overline{\rho^\prime}=0,\quad p=	\overline{p}+p^\prime, \quad \overline{p^\prime}=0, 
	\end{equation*}
	and the mass averaged fluctuations of the velocity, 
		\begin{equation*}
	u_i=	U_i+u_i^{\prime\prime},\quad  \overline{\rho u_i^{\prime\prime}}=0, 
	\end{equation*}
	The equations \eqref{System0} can be written as :
	\begin{equation}
	\left\{ 
	\begin{array}{l}
	\displaystyle\frac{\partial \overline{\rho}}{\partial t}+(\overline{\rho} U_i),_{i}=0,\\
	\\
		\displaystyle\frac{\partial \overline{\rho} U_i}{\partial t}+\left( \overline{\rho} U_i U_j +\overline{p} \,\delta
	_{ij}+R_{ij}\right),_{j}=0, \\
	\\
	\displaystyle \frac{dR_{is}}{dt}+R_{is}{U_{k,k}}+R_{ks}U_{i,k}+R_{ki}U_{s,k}=S_{is},
	\end{array}
		\right.
			\label{System1}
	\end{equation}
		where 
\begin{equation*}
{\mathbf R}=\{R_{is}\}, \quad R_{is}=\overline{\rho u_i^{\prime\prime}u_s^{\prime\prime}}, \quad (i,s=1,2,3), 
\end{equation*}
		and 
	\begin{equation*}
{\mathbf S}=\{S_{is}\}, \quad S_{is}=-\left(\overline{\rho u_i^{\prime\prime}u_s^{\prime\prime}u_k^{\prime\prime}}\right)_{,k}-\overline{p,_i u_s^{\prime \prime}}-\overline{p,_s u_i^{\prime\prime}}.
\end{equation*}
The expression of $S_{is}$ can also be written as
	\begin{equation*}
S_{is}=-\left(\overline{\rho u_i^{\prime\prime}u_s^{\prime\prime}u_k^{\prime\prime}}\right)_{,k}-\overline{h^{\prime},_i\rho u_s^{\prime \prime}}-\overline{h^\prime,_s \rho u_i^{\prime\prime}}, 
\end{equation*}
where $h(\rho)$ is the specific gas enthalpy : $$\displaystyle \frac{dh}{d\rho}=\frac{1}{\rho}\frac{dp}{d\rho}$$ and  $h^\prime$ means the enthalpy fluctuation.   

	System (\ref{System1}) can  be rewritten in  tensorial form as :\\
	\begin{equation*}
		\left\{
		\begin{array}{l}
			\displaystyle\frac{\partial \overline{\rho}}{\partial t}+%
			\mathrm{div}\left( \overline{\rho} \mathbf{U}\right) =0, \\
			\\
			\displaystyle\frac{\partial (\overline{\rho} {\mathbf{U}})}{\partial t}+\left[\mathrm{div}\Big( \overline{\rho} {\mathbf{U}}\otimes {\mathbf{U}}+\overline{p}\,\mathbf{I}+ {\mathbf{R}}\Big)\right]^T =0,
			\\ \\
			\displaystyle\frac{d\mathbf{R}}{dt}+\mathbf{R}\,\mathrm{div}\mathbf{U}%
			+\frac{\partial \mathbf{U}}{\partial \mathbf{x}}\,\mathbf{R}+\mathbf{R}%
			\left( \frac{\partial \mathbf{U}}{\partial \mathbf{x}}\right) ^{T}=\mathbf{S}%
			.%
		\end{array}%
		\right.   \label{System2}
	\end{equation*}%
 Using the mass
	conservation law, the equation of   volumic Reynolds stress tensor $\mathbf{R}$
	can be rewritten  for  the specific  Reynolds stress tensor $\mathbf{P}$ :
	\begin{equation*}
		\frac{d\mathbf{P}}{dt}+\frac{\partial \mathbf{U}}{\partial \mathbf{x}}\
		\mathbf{P}+\mathbf{P}\left( \frac{\partial \mathbf{U}}{\partial \mathbf{x}}%
		\right) ^{T}=\frac{\mathbf{S}}{\overline{\rho} } 
		\qquad \text{where}\quad
		\mathbf{P}=\frac{\mathbf{R}}{\overline{\rho} }.\label{Equation_3}
	\end{equation*}
	We  focus on the governing equations of mass conservation,  momentum equation and specific Reynolds stress evolution  without  source term $\mathbf S$. Such an ideal system (called sometimes \textit{ideal turbulence system}  \cite{Troshkin_1990}) appears as a natural step in applying a splitting-up technique in the numerical treatment of the full system of  compressible turbulence \cite{Audebert_2008,Berthon_2002}.
	\begin{equation}
		\left\{
		\begin{array}{l}
			\displaystyle\frac{\partial \overline{\rho} }{\partial t}+%
			\mathrm{div}\left( \overline{\rho} \mathbf{U}\right) =0, \\
			\\
			\displaystyle\frac{\partial (\overline{\rho} {\mathbf{U}})}{\partial t}+\left[\mathrm{div}\Big( \overline{\rho}  {\mathbf{U}}\otimes {\mathbf{U}}+\overline{p}\,\mathbf{I}+\overline{\rho}\,  {\mathbf{P}}\Big)\right]^T =0,\\
			\\
			\displaystyle \frac{d\mathbf{P}}{dt}+\frac{\partial \mathbf{U}}{\partial \mathbf{x}}\
			\mathbf{P}+\mathbf{P}\left( \frac{\partial \mathbf{U}}{\partial \mathbf{x}}%
			\right) ^{T}=0.
		\end{array}%
		\right.   \label{basic_equations}
	\end{equation}
	
A natural hypothesis is that the average pressure is a function of the averaged density (as for example, in  case of   isothermic ideal gas), \textit{i.e.} $\overline{p}=\mathcal{P}(\overline{\rho})$, and  we have a closed system \eqref{basic_equations}. \\	System (\ref{basic_equations}) admits the energy conservation
	law :
	\begin{eqnarray}
		\label{energy}
		&&\frac{\partial }{\partial t}\left(\overline{\rho}\left( \frac{1}{2}{\left\vert \mathbf{U}\right\vert ^{2}}+ \mathcal{E}+\frac{\mathrm{tr}\,\mathbf{P}}{2}\right) \right)\\ &&+\mathbf{\mathrm{div}}%
		\left( \overline{\rho}\mathbf{U}\left( \frac{1}{2}{\left\vert \mathbf{U}\right\vert ^{2}}+\mathcal{E}+\frac{\mathrm{tr}\,\mathbf{P}}{2}\right) +\overline{p}\mathbf{U}+\overline{\rho}\mathbf{P}\mathbf{U}%
		\right) =0, \notag
	\end{eqnarray}
	where
	$\mathcal{E}$ is the specific internal energy satisfying the Gibbs identity :
	\begin{equation*}
		d\mathcal{E}(\overline{\rho}) = \frac{\mathcal{P}(\overline{\rho})}{\overline{\rho}^2}\ d\overline{\rho}.
	\end{equation*}
	As proven in \cite{Gavrilyuk_Gouin_2012}, the last conservation law can be written in the form :
	\begin{equation*}
		\frac{\partial }{\partial t}\left(\overline{\rho}\Psi\right) +\mathbf{\mathrm{div}}\left(\overline{\rho}\Psi\mathbf{U}
		\right) =0,\quad\text{with}\quad \Psi =\frac{{\rm det}\mathbf{P}}{\overline{\rho}^2}.
	\end{equation*}
  System \eqref{basic_equations} coincides with the  asymptotic $2D$--model of weakly sheared shallow water flows \eqref{Teshukov_system}  when $\overline{\rho}$ is replaced by the fluid depth $h$, $\mathcal{P}(\overline{\rho})$ by $\displaystyle {gh^{2}}/{2}$,  and the specific energy    $\mathcal{E}(\overline{\rho})$ by $gh/2$.

	\subsection{Properties of the two models}
%
	\noindent 	In   particular case $\mathrm{{rot}\,}\mathbf{U}=\boldsymbol{0}$ (i.e. $\left( \partial \mathbf{U}/\partial \mathbf{x}
	\right) ^{T}=\partial \mathbf{U}/\partial \mathbf{x}$), the equation for the specific Reynolds stress tensor can be integrated 
   \cite 
	{Gouin}.  
Equation \eqref{basic_equations}$_3$ corresponds to the evolution of a two-covariant tensor convected
	by the mean flow. This means that $\mathbf{P}$ has a zero Lie derivative $d_{L} $ with respect to the  average velocity  $\mathbf{U}$. The solution of the equation  
	$$
	d_{L}\mathbf{P}\equiv \frac{d\mathbf{P}}{dt}+\frac{\partial \mathbf{U}}{%
		\partial \mathbf{x}}\ \mathbf{P}+\mathbf{P}\frac{\partial \mathbf{U}}{%
		\partial \mathbf{x}}=0,
	$$
	is
	\begin{equation*} 
		\mathbf{P}=\left( \mathbf{F}^{T}\right) ^{-1}\mathbf{P}%
		_{0}\left( \mathbf{X}\right)\mathbf{F}^{-1},
	\end{equation*}
	where $\mathbf{F}=\partial \mathbf{x/}\partial \mathbf{X}$ is the
	deformation gradient of the mean motion and   tensor $\mathbf{P}_0$ is the image of $\mathbf P$ in  Lagrange coordinates $\mathbf{X}$  \cite{Gouin,Serrin}.
	
	However,  hypothesis $\mathrm{{rot}\,}\mathbf{U}=0$ is not compatible with  systems  \eqref{Teshukov_system} and \eqref%
	{basic_equations}. Indeed, if initially $\mathrm{{rot}\,}\mathbf{U}=0$, this zero value   is not  conserved along the motion (the Kelvin theorem is not valid for system  \eqref{Teshukov_system} and system \eqref{basic_equations}), and the equation for $\mathbf{P}$ cannot be integrated in   Lagrange coordinates. The general case ($\mathrm{{rot}\,}\mathbf{U}\neq 0$) was studied in  \cite{Gavrilyuk_Gouin_2012}. \\
	
	Now we  ask  the following question : \\ Since   systems \eqref{Teshukov_system} and \eqref{basic_equations} are conservative, are we able to derive the governing equations from the Hamilton principle of stationary action as in      case of  classical non-dissipative models?\\    Both,   energy equations \eqref{energy1} and    \eqref{energy} suggest  to formulate Hamilton's action in the form :
	\begin{equation*}
		a=\int_{t_1}^{t_2}\mathcal{L}\, dt 
	\end{equation*}
	where for \eqref{energy1}, the Lagrangian is  
	\begin{equation*}
		\mathcal{L}=\int_{\Omega(t)} h\left( \frac{1}{2}{ \left\vert \mathbf{U}\right\vert ^{2}}-\frac{gh}{2}-\frac{\mathrm{tr}\,\mathbf{P }}{2} 
		\right) \,d\Omega ,
	\end{equation*}
and for \eqref{energy}, 
	\begin{equation*}
		\mathcal{L}=\int_{\Omega(t)} \overline{\rho}\left( \frac{1}{2}{ \left\vert \mathbf{U}\right\vert ^{2}}-\mathcal{E} \left(\overline{\rho}\right)-\frac{\mathrm{tr}\,\mathbf{P }}{2} 
		\right) \,d\Omega,
	\end{equation*}
where   $t_1$ and $t_2$\  are two fixed times, $\Omega(t)$ is the material volume associated with   average velocity $\mathbf U$ and $d\Omega$ denotes the  convected volume element in $\Omega(t)$. We have to set  which equations can be considered as imposed constraints and which equations are derived from Hamilton's principle.\\

	\section{Variational formulation for shear shallow water flows and ideal turbulence}
	\subsection{Virtual motion} 
	We recall the notion of {\it virtual motion} and of {\it virtual displacement}.   Let a one-parameter family of
	\emph{virtual motions} :
	\begin{equation*}
		\mathbf{x}=\mathbf{\Phi}\left(t, \mathbf{X},  \lambda \right)
		\label{vitual motion}
	\end{equation*}
	where $\mathbf{x}$ denotes the Euler  coordinates,  $\mathbf{X}$ denotes the Lagrange coordinates, $t$ is the time, and  $\lambda \in \cal{O}$ is a real number ($\cal{O}$ is an open interval containing $0$).  When $\lambda =0$,  
	$$\mathbf{\Phi}\left( t, \mathbf{X}, 0\right) =\boldsymbol{\phi }%
	\left(t,  \mathbf{X}\right), $$ where  $\boldsymbol{\phi }%
	\left(t,  \mathbf{X}\right) $  denotes the real motion associated with the averaged velocity field  $\mathbf U$.\\ As usually, we assume   at the boundary of $[t_1,t_2]\times \Omega(t)$,    $$\mathbf{\Phi}\left( t, \mathbf{X}, \lambda\right)=\boldsymbol{\phi }\left(t,  \mathbf{X}\right).$$ The virtual displacement of the particle is denoted  $\tilde\delta\mathbf{x}$ and is defined as   \cite{Serrin,Berdichevsky,Gavrilyuk_2011} :
	\begin{equation*}
		\tilde\delta\mathbf{x}(t, \mathbf{X})=\frac{\partial {\mathbf{\Phi}}(t,\mathbf{X}, \lambda)}{\partial \lambda}\vert_{\lambda=0}.
		\label{vitual displacement}
	\end{equation*}
	In the following, symbol $\tilde\delta$ means the derivative with respect to $\lambda$, at fixed Lagrange coordinates $\mathbf{X}$, when $\lambda=0$.
	We  denote by $\boldsymbol{\zeta}(t,{\mathbf x})$ the virtual displacement expressed as a function of  Euler  coordinates :
	\begin{equation*}
		\boldsymbol{\zeta}(t,\mathbf x)=\boldsymbol{\zeta}\left(t,{\boldsymbol \phi}(t, \mathbf X)\right)=\tilde\delta{\mathbf x}\left(t, \mathbf X\right).
	\end{equation*}
	As for $\boldsymbol{\zeta}$, for the sake  of simplicity,   we   use  for all   quantities the same notation in both Euler  and Lagrange coordinates.
	
	\subsection{Lagrangian}
	
	The equations for shear shallow water and ideal  turbulence  are identical when    we identify the  quantities $\displaystyle \overline{\rho} \; \text{and} \;  h, \;  \mathcal{E}\left(\overline{\rho}\right) \; \text{and} \; \frac{gh}{2}$.
	
	Consequently, let us consider the Lagrangian in the   general form :

	\begin{equation*}
		\mathcal{L}=\int_{\Omega(t)} \overline{\rho}\left( \frac{1}{2}{ \left\vert \mathbf{U}\right\vert ^{2}}-\mathcal{E}\left(\overline{\rho}\right)-\frac{\mathrm{tr}\,\mathbf{P }}{2} 
		\right) d\Omega\,.   
	\end{equation*}%
	We consider   two
	constraints :\\
	
	$\bullet$\quad The first one corresponds to the mass conservation law,
	\begin{equation*}
		\frac{\partial \overline{\rho} }{\partial t}+
		\mathrm{div}\left( \overline{\rho} \mathbf{U}\right) =0, 
	\end{equation*}
	which can be integrated in the form : 
	\begin{equation*}
	\overline{\rho}\, \det F =\rho_0(\mathbf X).
	\end{equation*}
	It corresponds to a holonomic constraint.\\
	
	$\bullet$\quad  The second one  corresponds to   the specific  Reynolds stress tensor evolution,
	\begin{equation*}
		\frac{d\mathbf{P}}{dt}+\frac{\partial \mathbf{U}}{\partial \mathbf{x}}\
		\mathbf{P}+\mathbf{P}\left( \frac{\partial \mathbf{U}}{\partial \mathbf{x}}%
		\right) ^{T}=0,
	\end{equation*}
	which is not integrable along the motion \cite{Gavrilyuk_2011}. So, it corresponds to a  non-holonomic constraint.  \\
	
	Two types of variations for unknowns $\overline{\rho}$, $\mathbf U$ and $\mathbf P$  can be used  \cite{Berdichevsky, Gavrilyuk_2011} :\\ 
	
	$\bullet$\quad The previous one, at fixed Lagrangian coordinates (denoted by $\tilde\delta$),  \\

	$\bullet$\quad Another equivalent variation at fixed Eulerian coordinates (denoted by $\hat\delta$).\\
	
	\noindent  These variations are related : for any variable $f$,  the connection between the two variations is :
	\begin{equation}
		\hat\delta f=\tilde\delta f-{\boldsymbol\nabla f}\cdot{\boldsymbol \zeta}.
		\label{relation}
	\end{equation}
	We consider that the gradient operator,  as all  space operators,  is taken in Euler  coordinates. The mass constraint  allows us 
	to obtain the variation of $\overline{\rho}$ at fixed Lagrange and Euler  coordinates  in the form   \cite{Berdichevsky, Gavrilyuk_2011} :
	\begin{equation}
		\tilde\delta \overline{\rho}=-\overline{\rho}\,\mathrm{div}(\boldsymbol{\zeta})\quad\text{and}\quad \hat{\delta} \overline{\rho}=-\mathrm{div}(\overline{\rho}\,\boldsymbol{\zeta}).\label{rhovar}
	\end{equation}
	The  variations  of velocity  $\mathbf{U}$  at fixed Lagrange (or Euler) coordinates are given respectively as  \cite{Berdichevsky, Gavrilyuk_2011} :
	\begin{equation}
		\tilde {\delta} \mathbf{U }= \frac{\partial \tilde\delta \mathbf{x}}{\partial t}=\frac{d \boldsymbol{\zeta}}{dt} \quad\text{and}\quad \hat{\delta} \mathbf{U }= \frac{d \boldsymbol{\zeta}}{dt}-\frac{\partial \mathbf{U}}{\partial \mathbf{x}}\boldsymbol{\zeta }.\label{Uvar}
	\end{equation}
	However,  equation \eqref{basic_equations}$_3$ for $\mathbf P$ is not integrable in  Lagrange coordinates.  Let us recall that  $m$  non-holonomic constraints in analytical mechanics for  a system with $n$  degrees of freedom $\mathbf q=(q_1, q_2,...,q_n)^T$, where  $m<n$, are in the form :
	\begin{equation*}
		{\mathbf A }({\mathbf q}, t)\,\frac{d\mathbf q}{dt} +{\mathbf b(t)}=0.
		\label{non-holonomic_constraint}
	\end{equation*}
	Matrix $\mathbf A$ is a matrix with $n$ columns and $m$ lines  and $\mathbf b$ is a   time dependent vector in $\mathbb{R}^n$. Even if the  system of constraints cannot be reduced  to pure  holonomic constraints,  the  variations of $\mathbf q$ corresponding to these non-holonomic constraints  are expressed as \cite{Zhuravlev} :
	\begin{equation*}
		{\mathbf A }({\mathbf q}, t)\,\delta {\mathbf q}=0.
	\end{equation*}
	Similarly,  equation for $\mathbf{P}$ can be seen  as a non-holonomic constraint, and consequently the Lagrangian variation of $\mathbf{P}$ can be written  in the form :
	\begin{equation*}
		\tilde\delta\mathbf{P}=-\frac{\partial \boldsymbol{\zeta}}{\partial \mathbf{%
				x}}\,\mathbf{P}-\mathbf{P}\left( \frac{\partial \boldsymbol{\zeta}}{\partial
			\mathbf{x}}\right) ^{T}.\label{key1}
	\end{equation*}
	It implies :
	\begin{equation*}
		\tilde \delta (\overline{\rho}\,\mathbf{P})=-\overline{\rho}\,\frac{\partial \boldsymbol{\zeta}}{\partial
			\mathbf{x}}\,\mathbf{P}-\overline{\rho}\,\mathbf{P}\left( \frac{\partial \boldsymbol{\zeta}}{\partial
			\mathbf{x}}\right) ^{T}-\overline{\rho}\,\mathbf{P}\,{\rm div}{\boldsymbol \zeta}.
	\end{equation*}
	Since the operator ${\rm tr}$  and variation $\tilde \delta$ commute,  we obtain :
	\begin{equation*}
		\tilde \delta \left[{\rm tr}(\overline{\rho}\,\mathbf{P})\right] =-2\,{\rm tr}\left(\overline{\rho}\,\mathbf{P}\,\frac{\partial \boldsymbol{\zeta}}{\partial
			\mathbf{x}}\right)-{\rm tr}(\overline{\rho}\,\mathbf{P})\ {\rm div}{\boldsymbol \zeta}.
	\end{equation*}
	Its Eulerian variation $\hat\delta$ (considered at fixed Euler  coordinates)  is obtained according to relation \eqref{relation}  :
	\begin{eqnarray}
		\hat\delta \left[{\rm tr}(\overline{\rho}\,\mathbf{P})\right]&=&-2\,{\rm tr}\left(\overline{\rho}\,\mathbf{P}\,\frac{\partial \boldsymbol{\zeta}}{\partial
			\mathbf{x}}\right)-{\rm tr}(\overline{\rho}\,\mathbf{P})\,{\rm div}{\boldsymbol \zeta}-\{\boldsymbol \nabla  \left[{\rm tr}(\overline{\rho}\,\mathbf{P})\right]\}^T\, \boldsymbol \zeta\notag
		\\
		&=&-2\,{\rm tr}\left(\overline{\rho}\,\mathbf{P}\,\frac{\partial \boldsymbol{\zeta}}{\partial
			\mathbf{x}}\right)-{\rm div}\left[{{\rm tr}(\overline{\rho}\,\mathbf{P})\boldsymbol \zeta}\right].\label{Pvar}
	\end{eqnarray}
	
	The variation of Hamilton's action in Euler  coordinates is :
	\begin{equation*}
		\hat\delta a=\int_{t_{1}}^{t_{2}}\int_{\Omega(t)}\left(\frac{\hat\delta \overline{\rho}}{2}\,{ \left\vert \mathbf{U}\right\vert^{2}}+\overline{\rho}\,\mathbf{U}^T\,\hat\delta\mathbf{U}-\frac{\partial (\overline{\rho}\,\mathcal{E}) }{\partial\overline{\rho}}\,\hat\delta\overline{\rho}-\frac{\hat\delta\, {\rm tr}(\overline{\rho}\,\mathbf{P})}{2}\right)
		d\Omega\, dt .
	\end{equation*}
	By using  formula \eqref{rhovar}, \eqref{Uvar} and \eqref{Pvar}  for   Euler  variations, we obtain :
	\begin{eqnarray*}
		\hat\delta a &=& \int_{t_{1}}^{t_{2}}\int_{\Omega(t)}\left[\ -\frac{\mathrm{div}
			(\overline{\rho}\,\boldsymbol{\zeta})}{2}\,{ \left\vert \mathbf{U}\right\vert^{2}}+\overline{\rho}\,\mathbf{U}^T\,\left(\frac{d \boldsymbol{\zeta}}{dt}-\frac{\partial \mathbf{U}}{\partial \mathbf{x}}\boldsymbol{\zeta }\right)\right. 
		\\  &+&\left.   \frac{\partial (\overline{\rho}\,\mathcal{E}) }{\partial\overline{\rho}}\, \mathrm{div}(\overline{\rho}\,\boldsymbol{\zeta}) + {\rm tr}\left(\overline{\rho}\,\mathbf{P}\,\displaystyle\frac{\partial \boldsymbol{\zeta}}{\partial
			\mathbf{x}}\right)+\frac{1}{2}\,{\rm div}\left[{\rm tr}(\overline{\rho}\,\mathbf{P}){\boldsymbol \zeta}\ \right]  \right]\,
		d\Omega\, dt .
	\end{eqnarray*}\\
	The Gauss-Ostrogradsky formula and the fact that the variations vanish at the boundary of the domain $[t_1,t_2]\times \Omega_t$ imply :
	\begin{equation*}
		\hat\delta a=-\int_{t_{1}}^{t_{2}}\int_{\Omega(t)} \left[\displaystyle\frac{ \partial (\overline{\rho}\, {\mathbf{U}})^T}{\partial t}+\mathrm{div} \Big( \overline{\rho}\,  {\mathbf{U}}\otimes {\mathbf{U}}+\mathcal{P}(\overline{\rho})\,\mathbf{I}+\overline{\rho} \, {\mathbf{P}}\Big)\,\right]  {\boldsymbol \zeta}\ d\Omega\,dt.
	\end{equation*}
For all vector field ${\boldsymbol \zeta}$, the variation of Hamilton's action in Euler coordinates vanishes, the fundamental lemma of variation calculus yields  momentum equation \eqref{basic_equations}$_2$.
	
\noindent	The case of non-isentropic compressible turbulent flows can be treated in the same way.

	\section{Conclusion}
	
	We have established that the momentum equation of the non-dissipative model for shear shallow water flows and for ideal turbulence  can be obtained by the Hamilton principle of stationary  action. As usually, the mass conservation law corresponds to a holonomic (or integrable) constraint, but the evolution equation for  the Reynolds stress tensor which is not integrable, corresponds to  a non-holonomic constraint. 
	
	Systems \eqref{Teshukov_system} and \eqref{basic_equations} belong to the class of physical models subject to Hamilton's principle of stationary action, as is generally the case for conservative systems with holonomic constraints. 
	

	Systems \eqref{Teshukov_system} and \eqref{basic_equations} are hyperbolic and shock waves can be formed. It can    be proved, that systems \eqref{Teshukov_system}   and \eqref{basic_equations} cannot be written in conservative form \cite{Gavrilyuk_2018}. So, a  definition of weak solutions is questionable. The fact that the equations admit the  variational formulation could  allow us  to formulate the corresponding Rankine-Hugoniot relations (shock relations) for shear shallow water flows  as it was done, for  example, for two-velocity flows in \cite{Gavrilyuk_Gouin_Perepechko_1998}.  They could confirm  empirical  Rankine--Hugoniot relations proposed for systems \eqref{Teshukov_system} and  \eqref{basic_equations} in \cite{Gavrilyuk_2018, Ivanova_2019}.

	\end{document}